# Validation of Sentinel-3A SARAL coastal sea level data at high posting rate: 80 Hz

A. Aldarias, J. Gómez-Enri, I. Laiz, B. Tejedor, S. Vignudelli, and P. Cipollini

*Abstract*— Two and a half years (June 2016 - November 2018) of Sentinel-3A SRAL (S3A-SRAL) altimetry data were validated at the sampling frequency of 80 Hz. They were obtained from the European Space Agency GPOD service over three coastal sites in Spain: Huelva (Gulf of Cadiz), Barcelona (Western Mediterranean Sea) and Bilbao (Bay of Biscay). Two tracks were selected in each site, one ascending and one descending. Data were validated using in-situ tide gauge data provided by the Spanish Puertos del Estado. The altimetry sea level anomaly time series were obtained using the corrections available in GPOD with the exception of the sea state bias correction (SSB), not available at 80 Hz. Hence, the SSB was approximated to 5% of the significant wave height. The validation was performed using two statistical parameters, the Pearson correlation coefficient (r) and the mean square error (rmse). In the 5-20 km segment with respect to the coastline, the results were 6-8 cm (rmse) and 0.7-0.8 (r) for all the tracks. The 0-5 km segment was also analysed in detail to study the land effect on the altimetry data quality. Results showed that the track orientation, the angle of intersection with the coast, and the land topography concur to determine the nearest distance to coast at which the data retain a similar level of accuracy than in the 5-20 km segment. This —distance of good quality‖ to shore reaches a minimum of 3 km for the tracks at Huelva and the descending track at Barcelona.

## I. INTRODUCTION

MORE than 25 years of satellite radar altimetry have fully demonstrated its value to monitor global sea level [1]. A great challenge being faced by the altimetry community is to improve the quality of data in coastal areas [2], leading to reliable estimates of regional and local sea level. The coastal altimetry community has faced two problems near the shore. On one hand, there is a high level of uncertainty associated with the estimation of the geophysical parameters (retracked range, significant wave height, and wind speed at the sea surface) derived from the processing of the radar return signal (retracking), which is mainly due to the land contamination of the radar waveforms [3]. On the other hand, the quality of some range and geophysical corrections used to estimate the sea level (mainly wet tropospheric, tidal, and sea state bias corrections) needs improving [4], [5], [6], [7], [8], [9], [10], [11], [12], [13], [14], [15], [16], [17], [18], [19].

While products from pulse-limited altimetry have traditionally been provided and used at a posting rate of 1 Hz (which corresponds to an along-track spatial resolution of 7 km between two consecutive measurements), higher rate products (10, 18, 20, or 40 Hz depending on the particular mission, equivalent to 700, 388, 350, and 175 m of along-track spatial resolution, respectively) are also available. Some authors compared products at different resolutions in coastal areas in order to assess the quality of these data. A validation study of Sea Level Anomaly (SLA) was performed in the closest 30 km to the north-western coast of the Mediterranean Sea using in-situ tide gauge measurements [20]. They compared different altimetry product of Topex/Poseidon at 1 Hz and 10 Hz, and Jason-1 and Jason-2 at 1 Hz and 20 Hz. In the case of Jason-1, the correlation coefficients (r) were 0.70 – 0.76 and 0.73 – 0.80 at 1 Hz and 20 Hz posting rate, respectively. For Jason-2, the r values were 0.69 – 0.79 (1 Hz) and 0.72 – 0.82 (20 Hz). The results indicated an increase in the quality and in the number of data available at 20 Hz in coastal areas.

Then the pulse-limited satellite SARAL/AltiKa, designed to analyse mesoscale processes, was launched in 2013. This operates in the Ka-band with the advantage of a smaller footprint [21], [22]. SARAL/AltiKa SLA data at 20 Hz were validated in Huelva (Gulf of Cadiz, GoC hereinafter) and it was observed that the rmse values increased toward the coast when compared with in-situ data [10]. For example, the authors obtained root mean square error (rmse) values of 5.3 cm at 10 km, increasing to 35.4 cm at 5 km and to 57.0 cm at 1 km. This clearly highlights the need for improving the quality of coastal altimetry data. In a previous study, SARAL/AltiKa SLA time series at 40 Hz in the Strait of Gibraltar had been compared with in-situ measurements obtaining along-track rmse values between 8 and 10 cm at 7 km from the coast [23].

Further advances in coastal altimetry are due to the Synthetic Aperture Radar (SAR) mode of CryoSat-2 and now on Sentinel-3A and -3B. The improvements are due to a smaller uncertainty associated with the estimation of the geophysical parameters. SAR-mode altimetry has a smaller footprint in the along-track direction with respect to the pulse-limited footprint, which improves the along-track spatial resolution, reduces the noise, and might decrease the land contamination

A. Aldarias, J. Gómez-Enri, I. Laiz and B. Tejedor are with the department of Applied Physics, University of Cadiz, Puerto Real, Spain (e-mail:{anaisabel.aldarias, jesus.gomez, irene.laiz, begonia.tejedor} @uca.es).

S. Vignudelli is with the Institute of Biophysics, National Research Council. Pisa, Italy. (e-mail: vignudelli@pi.ibf.cnr.it).

P. Cipollini is with Telespazio VEGA UK Ltd. for ESA Climate Office, ESA-ECSAT, Harwell Campus, Didcot, U.K. (e-mail: paolo.cipollini@esa.int)

of the radar waveforms [8], [12]. Therefore, it allows the study of shorter scale ocean features and processes. Another advantage is a higher quality of the parameters derived from the altimeter measurements (see [24], [8], [14] among others).

CryoSat-2 SLA data at 20 Hz were validated using tide gauges in the German Bight and West Baltic Sea, for the 0 – 10 km segment from the coast [14]. The study evidenced improvements from the SAR mode with respect to the PLRM (Pseudo Low Resolution Mode) within the coastal zone. The authors calculated the standard ~~derivation~~ deviation of the difference (stdd) and the correlation coefficient, obtaining average values of 4.4 cm and 0.96, respectively. A similar study [10] was carried out in Huelva for the 5 – 20 km segment, obtaining rmse values of 6.4 cm at 20 km, increasing to 8.5 cm at 5 km and 29.3 cm at 1 km.

The objective of the present paper is to validate time series of SLA from Sentinel-3A SRAL (SAR Radar ALtimeter) (S3A-SRAL, hereinafter) close to the coast, using in-situ tide gauge measurements along three locations of the Spanish coast. What makes this study novel is that S3A-SRAL sea level data is analysed at the highest posting rate available, 80 Hz, equivalent to a distance between two consecutive along-track measurements of about 85 m. The number of valid data near the coast at the three locations, the orientation of the satellite tracks with respect to the land intersection, and the land contamination of the radar waveforms in the 0 – 5 km track segments is also analysed.

The paper is organised as follows: the study areas are presented in Section 2. The data sets and the methodology are presented in Section 3 and 4, respectively. Section 5 and 6 show the results and discussion, followed by the final conclusions in Section 7.

## II. STUDY AREAS

The selection of the study sites is made based on the availability of radar tide gauge data in regions where both an ascending and a descending track passed close to the in-situ instrument's location.

The validation of S3A-SRAL along-track sea level data is made using three tide gauges located in three coastal zones of the Iberian Peninsula: Huelva (HU_TG), Barcelona (BA_TG), and Bilbao (BI_TG). The three areas are characterised by coastal environments with different tidal and hydrodynamic conditions. Six S3A-SRAL tracks were selected considering their proximity to the tide gauges. All the tracks pass closer than 35 km to the ground truth instruments (Table I) (Figure 1).

The HU_TG is located in the eastern shelf of the GoC, Southwest of the Iberian Peninsula. Tides are mainly mesotidal, with amplitudes above 1 m. The GoC surface circulation is characterised by a strong seasonality that is linked to the offshore circulation: some authors proposed the existence of a mesoscale cyclonic cell over the eastern continental shelf during spring-summer [25], [26], its northern part being a warm coastal countercurrent [27], [28]. This countercurrent is generally replaced by an eastward flowing current during autumn and winter [29]. More recent studies suggest that the onset of this countercurrent is a common feature over the year and does not show a seasonal behaviour [30]. The Guadalquivir River also plays an important role in the eastern GoC surface circulation. Sporadic but heavy freshwater discharges might contribute to the sea level at different time-scales [31], [9], [10]. The study area has also been used in the past for the validation of conventional pulse-limited and SAR-mode radar altimeter data near the shore. Some authors used weekly gridded maps of SLA from AVISO (Archiving, Validation and Interpretation of Satellite data in Oceanography) to validate SLA time series at different time scales finding high and significant correlations (r > 0.85) with in-situ tide gauge sea level data at monthly time scales [31], [32]. As mentioned before, SLA time series from the CryoSat-2 SIRAL altimeter, in SAR mode, were validated using the HU_TG tide gauge [10]. The authors analysed along-track data with an along- track spatial resolution of 20 Hz, obtaining rmse of 6.4 – 8.5 cm in the 5 – 20 km segment from the coast.

The BA_TG is located in the northeastern part of the Iberian Peninsula, within the western Mediterranean Sea. The range of tidal amplitudes into the Mediterranean Sea is smaller than 1 m (i.e. microtidal). The surface circulation in the northwestern part of the Mediterranean Sea is generally cyclonic and thermohaline [33]. The current in this zone (Liguro-Provençal Current or LPC) follows a south-southwest direction. This current is determined by the Coriolis force and the pressure gradient between the less salty waters (shelf waters) and the denser water masses (continental slope) [34]. The coastal hydrodynamics is mainly controlled by the local meteorology, the oceanography and the submarine physiography. The main river tributaries are Rhone and Ebro, two of the largest Mediterranean rivers [35].

The BI_TG is situated in an embayment within the eastern North Atlantic Ocean, the Bay of Biscay. The Spanish shelf in this area is narrow (30 to 40 km) in comparison with the French shelf [36]. The Bay of Biscay is a region of large tidal amplitudes, between 1-4 m, of mesotidal type [37], and a strong thermohaline forcing [38]. The circulation system is dominated by a weak anticyclonic circulation in spring and summer [39], a slope current, the input of freshwaters by various rivers, coastal upwellings, the northward flow of Mediterranean water, the transport along the submarine canyons [40], and the presence of mesoscale features, such as cyclonic and anticyclonic eddies. These features affect the exchange between the abyssal plain and the shelf [39], [41], [42]. Rivers in the Cantabrian shelf are short due to the proximity of the mountains to the sea and their total run-off represents a third of the sum of the Loire and Gironde, in the French margin. The regional ocean model MARS3D (Model for Application at Regional Scale) has been developed to study the hydrology of the Bay of Biscay reproducing the dynamics of the main river plumes over the shelf [43], [44]. Density currents are observed in the vicinity of estuaries due to the river freshwater discharges; these induce a poleward circulation modulated by wind [45], [46]. The maximum flow rates are mainly in spring and the minimum rates at the end of summer [47]. The main wind direction in the Bay of Biscay shifts from south-westerly to north-westerly [48].

## III. DATA SETS

### A. Altimeter data

Sentinel-3A is part of the Sentinels constellation of the Copernicus program. This program is developed by the European Commission (EC) in collaboration with the European Space Agency (ESA) and the European Organisation for the Exploitation of Meteorological Satellites (EUMETSAT) [49]. Data from S3A-SRAL were provided by the ESA Grid Processing On Demand (GPOD) SARvatore (SAR versatile altimetric toolkit for ocean research & exploitation) service, available at: https://gpod.eo.esa.int/. SARvatore-GPOD is based on the processing of S3A-SRAL data from Level 1a data or Full Bit Rate (FBR), up to Level 2. The SARvatore-GPOD service allows the user to configure the data processing at two processing levels: Level 1b, in which the multilooked radar waveforms are built, and Level 2 allowing the retrieval of the geophysical parameters. The S3A-SRAL data processing was restricted to the study areas using a GPOD processing baseline tailored for coastal zones. The GPOD options selected for processing the data for this study at Level 1b were: Hamming weighting, window in azimuth, approximated Doppler Beam steering (beam-forming), FFT zero-padding, and wider stack subsetting to give a final radar receiving window size of 512 samples; at Level 2, the options were: restricting the retracking to specific surface in all passes and applying a PTR width alpha parameter obtained from a Look-Up-Table. The retracking method was SAMOSA+ (SAR Altimetry MOde Studies and Applications), which is the SAMOSA 2 tailored for application in open ocean, coastal zone and ice [50]. Dinardo provided a complete description of this processing, and its impact on the data output [51].

S3A-SRAL is in orbit since February 2016 with a track revisit time of 27 days. Data at GPOD are available since June 2016. Six ascending/descending tracks have been analysed (Table I). The time period selected spanned from June/July 2016 (cycle 5 or 6, depending on the track) to October 2018 (cycle 36 or 37). A total of 32 cycles were retrieved in all cases. The along-track sampling rate selected was the highest of this altimeter in SAR mode: 80 Hz, equivalent to a distance between two consecutive along-track measurements of about 85 m.

### B. In-situ tide gauge data

Water levels from the Spanish Puertos del Estado tide gauge network (http://www.puertos.es) were used. This network is part of the Red de Mareógrafos (REDMAR), which is integrated in the Permanent Service for Mean Sea Level (PSMSL) and the Global Sea Level Observing System (GLOSS). The instruments are located at the ports of Huelva, Bilbao and Barcelona (Figure 1). They are all radars, measuring with a frequency of 2 Hz. Although the real time product is available at a 1-minute interval, Puertos del Estado also provides a 5-minute delayed product that has passed a standard quality control [52].

### C. Wave data

Hourly wave height and period data were provided by Puertos del Estado, already corrected using their standard quality control procedures [53], [54], [55]. The wave buoys used in this study were: The GoC deep buoy, located at 69 km from HU_TG, the Barcelona coastal buoy, located at 4 km from BA_TG, and the Bilbao-Vizcaya deep buoy, located at 32 km of BI_TG.

### D. Land Topography data

Land topography was obtained from the Shuttle Radar Topography Mission (SRTM). The product used here was available at http://srtm.csi.cgiar.org/. This is version 4 of the 90 m spatial resolution product [56]. The product from CGIAR-CSI (Consortium for Spatial Information) is based on the finished-grade 3 arc-second SRTM data processed by NASA.

## IV. METHODOLOGY

Concomitant time series of SLA from altimeter and tide gauge data were obtained at the three locations.

### A. Altimetry

The time series at 80 Hz posting rate were built in four steps: (1) The latitude-longitude positions of the tracks intersection with the coast were obtained using the Google Earth's .kmz files available in GPOD; (2) the 0 – 20 km track segments were selected using the positions obtained in step (1); (3) the distance between two consecutive radar measurements was checked within each cycle to ensure that it remained constant (~85 m); and (4) the time series were constructed using the 32 cycles along the closest tracks to the tide gauges. The maximum distance between the track segments and the tide gauges was 32.4 km (Table I). The SLA time series were calculated following equation (1):

$$\tag{1}$$

where *Orbit* (or Altitude) is the distance between the satellite's centre of mass and the reference surface (ellipsoid WGS84). *Range* is the retracked distance between the instrument and the mean reflected surface [8], [45] and it was obtained using the SAMOSA+ retracker [45]. *Range corrections* include the dry and wet tropospheric effect from the European Centre for Medium-range Weather Forecasts (ECMWF) models, and the ionospheric correction from the Global Ionospheric Maps (GIM) of the Jet Propulsion Laboratory. Although there are various alternatives for the wet tropospheric correction [15], [19], [57], GPOD only provides the one based on the ECMWF model. The *Geophysical corrections* include the ocean equilibrium tide, the ocean long period, the ocean load tide, the solid earth tide, and the pole tide. The atmospheric effects included in the Dynamic Atmospheric Correction (DAC) were not applied to Eq. (1). The tidal correction was obtained from the TPX08 (Topex Poseidon Global Inverse Solution version 8) tide model [58], whose accuracy is analysed in section 5.1. The Sea State Bias (SSB) correction is not available in the GPOD service at the posting rate used here (80 Hz). Section 5.2 is devoted to the analysis of a solution to overcome the lack of this correction, based on the Significant Wave Height (SWH) available in the product. The *Mean Sea Surface* (MSS) used was DTU15 [59], [60], the only one available in GPOD at the time of downloading the data.

### B. Tide gauge

The 5-minute delayed tide gauge data were used in this study. The time series were built using the time of the altimeter data. The temporal difference between altimeter and tide gauge data was below 2.5 minutes. The *Sea Levels* were obtained following equation (2):

$$\text{Water Level} - \text{Tide Prediction} \qquad (2)$$

where *Water Level* is the sea level measurement and the *Tide Prediction* was calculated from the tide gauges data with a classical harmonic analysis [61].

### C. Comparison between time series

Firstly, the temporal mean of the time series was eliminated to obtain the anomalies: S3A_SLA and TG_SLA, for the altimeter and tide gauge data, respectively. In a second step, a data screening was used to remove outliers: (i) the values outside the range: [-1.5 1.5] m; and (ii) the values outside the median ± 3$\sigma$. Finally, an extra outlier detector was applied to the time series of altimeter measurements based on the along-track SWH (available at 80-Hz along the track segments). All the measurements outside the range: [0 – 8] m were considered as an invalid measurement for the computation of the SSB and thus, discarded. About 5% of the along-track SWH were negative and about 1% bigger than 8 m. The selection of 8 m was made based on the analysis of the wave height roses from the closest buoys to the tide gauge stations for the period 2016-2018 (see Figure 2). As shown in Figure 2, SWH > 8 m are not observed in the study areas along the analysed time period. Only those tracks with at least 20% of valid cycles, i.e., cycles that successfully passed these screening, were considered for the validation.

The S3A_SRAL time series were validated with the TG_SLA ones using two statistical parameters, namely, the r coefficient and the rmse, as in previous works [50], [8], [9], [10], [12], [14]. The validation focused on the along-track segment located between 0 and 20 km, with 0 km corresponding to the point where the track intersects the coast.

## V. RESULTS

### A. Tidal model assessment: TPXO8

The TPX08 tide model generates a basic global solution for the main harmonic constituents with 1/30-degree resolution in the majority of the coastal areas (and 1/6-degree resolution in the rest of the ocean). The tide model was assessed by comparing the constituents available in the model with the constituents obtained from the tide gauge data analysis. The comparison was made with the closest tidal points to the tide gauges (Figure 1). The amplitudes and phases of the 9 constituents available in the global model were compared with those obtained with the tide gauges, and are shown in Table II. The assessment was performed by estimating the root mean squares (rms) and the residual sum of squares (rss) between the tidal model constituents and the tide gauges data [62]. A summary of the results is shown in Table III. The rms is smaller than 2 cm for all the constituents and locations with the exception of the M2 at Bilbao (4.99 cm). The rss is below 2 cm at Huelva and Barcelona, and up to 5 cm at Bilbao. This could be due to the fact that the tidal point used at Bilbao is about 6 km away from the tide gauge. The difference in amplitude and phase of the M2 between the model and the tide gauge (Table II) at this location explains the magnitude of the rms and rss, respectively. The results at Huelva are in agreement with a previous work [10], where an rss of 2.4 cm was obtained using the outputs of the DTU10 tidal model [63] and the same tide gauges used here. Hence, the TPX08 model seems to be a good choice to reproduce the tides at the three locations.

### B. Sea State Bias correction (SSB)

As mentioned before, SSB correction is not available in the GPOD service at 80-Hz posting rate. In order to get a first approximation of this correction, a parametric approach was made by estimating the rmse between S3SA_SLA and TG_SLA, when SSB values ranging between 0% (no correction) and 10% of the SWH ~~where~~ were applied in Eq. (1). The mean rmse for the track segment between 5 and 20 km to the land intersection was estimated. The reason why this track segment was selected is explained in sections 5.3 and 5.4. The results are shown in Figure 3 for Huelva (Figure 3.a), Barcelona (Figure 3.b), and Bilbao (Figure 3.c), respectively. Overall, the smaller rmse values are observed for SSB corrections ranging between 4% and 7% of the SWH, with very small differences among them of about ± 0.01 cm. Hence, a SSB correction of 5% of the SWH was used at all the locations and tracks, in agreement with previous works [8], [10].

### C. Availability of altimeter data near the coast

The number of valid data along the track segments are analysed in detail taking into account their proximity to coastal crossing, measured as the distance from the intersection of the track with the closest land. Figure 4 shows the number of valid data along the whole track segment, defined as a distance of 0 - 20 km to the tracks intersection point with the land. The maximum number of valid data is 32 (corresponding to 32 27-day cycles) for all the tracks. Note that three cycles of track #114 (Huelva) were considered invalid due to the lack of the ocean equilibrium tide correction. The number of valid data at Huelva is 32 (#322) / 29 (#114) up to 2 km to the land intersection (Figure 4.a). The same is observed for track #008 at Barcelona (Figure 4.b) and #051 at Bilbao (Figure 4.c). Track #356 (Barcelona) shows a strong reduction in the number of valid data in the $0 – 5$ segment. Finally, track #071 (Bilbao) shows around 20 valid data in the track segment $5 – 20$ km, and a considerably smaller number along the $0 – 5$ km segment.

Table IV summarises the comparison between the S3A_SLA and TG_SLA time series in the $0 – 5$ km, $5 – 20$ km, and $0 – 20$ km segments at the three locations. The mean rmse, r, confidence level (CL) and percentage of valid data (VD) were calculated in the three considered track segments. The smaller rmse and larger r values correspond to the $5 – 20$ km segment for all the tracks and locations. The $0 – 5$ km segment, however, shows the largest rmse and smaller r for all the locations and tracks. No valid numbers are shown on track #071 within this segment due to the low percentage of valid data (15.6%) obtained. These results reinforce the idea of avoiding using the closest track segments to the coast. A more detailed analysis of the $0 – 5$ km segment is given in section 6.

### D. Validation of along-track S3A-SRAL data

Along-track r (Figure 5) and rmse (Figure 6) values are shown at the three locations at 80 Hz posting rate in the $5 – 20$ km segment. Overall, larger correlations correspond to smaller rmse for all the locations and tracks. The $5 – 7$ km segment shows slightly smaller / bigger r / rmse in three tracks: #322 (Huelva), #356 (Barcelona), #051, and #071 (Bilbao). Good correlation (> 0.8, 95% C.L.) and small rmse (<8 cm) values are obtained along the $7 – 20$ km segment for these tracks and along the $5 – 20$ km segment for tracks: #114 (Huelva), and #008 (Barcelona). The only exception is observed on track #071 (Bilbao) (Figure 5.c and Figure 6.c) at distances between 12 and 17 km to the land intersection. In order to explain this, a more detailed analysis was carried out. The cycle-by-cycle along-track S3A_SLA is shown in Figure 7.a. Missing data (i.e., non-valid measurements) are observed along most of the track segments in a few cycles as a result of the data screening process. Cycle 32 (June 5, 2018) shows anomalous (negative) values of SLA in the $12 – 17$ km segment (Figure 7.b). A deeper investigation of the corrections used in Eq. (1) shows that these values correspond to strong negative values in the SSB, which are due to large SWH values. However, the in-situ wave buoy time series for that date does not show such high SWH values (Figure 2.c), the maximum wave heights being about 1.5 m. Thus, retracked high waves could be due to the retracking failing to accurately retrieve the SWH. Figure 8.b shows the radargram of the waveforms (cycle 32). Note that only the power from gates 300 to 400 is shown. High power is observed in the leading edge area at distances from the coast bigger than 17 km, but in the $12 – 17$ km segment the power has a much lower level, while higher than average power levels appear in the trailing edge portion of the waveform. A possible explanation for this distribution of power in the waveforms could be the presence of downslope wind gusts in the coastal zone, combined with calmer conditions offshore, with a transition zone in the 12 -17 km distance. Such a distribution of power complicates the retracking of the waveforms and hence the retrieval of accurate geophysical parameters, especially SWH. Hence, the same analysis shown in Figures 5 and 6 was performed for track #071, but excluding cycle 32 from the time series before the comparison with in-situ data. The removal of cycle 32 does not affect the along-track values of r (Figure 9.a) and rmse (Figure 9.b), with the exception of the $12 – 17$ km segment, where better results are clearly observed. The removal of cycle 32 reduces the average rmse in the $5 – 20$ track segment from 8.2 cm to 7.3 cm. This example highlights the importance of designing an accurate data screening strategy, supported with in situ data.

A similar, but significantly weaker behaviour can be found on track #008 (Barcelona) in the $7 – 14$ km segment (Figures 5 and 6). The analysis of the cycle-by-cycle along-track S3A_SLA (not shown here) presents anomalous values of SLA on cycle 06 (June 29, 2016), also related to high SWH values. The removal of cycle 6 from the S3A_SLA time series presents better results in terms of r (Figure 10.a) and rmse (Figure 10.b) along that track segment. Removing cycle 06 from the analysis the average rmse drops to 5.4 cm in the $5 – 20$ km track segment. Finally, the same problem was observed on track #322 (Huelva). In this case (not shown here), cycle 11 (December 3, 2016) showed high waves in the $5 – 6$ km track segment not supported by the in- situ wave height data. Removing that cycle from the analysis dropped the average rmse to 5.4 cm.

### VI. DISCUSSION

The coastal altimetry community is focused on the retrieval of accurate measurements near the coast. The results obtained in this work demonstrate that the S3A-SRAL altimeter gives accurate sea level data in the $5 – 20$ km coastal ocean fringe at the highest posting rate available as a standard product, i.e., 80 Hz. The values of r and rmse obtained in this study are between 0.7-0.8 and 6-8 cm, respectively, at all the locations. They are in line with previous validation works using other altimeters in coastal areas. A study in the northwestern Mediterranean Sea obtained correlation coefficients of 0.7-0.8 along the $0 - 30$ km segment for Topex/Poseidon (10 Hz), and Jason 1/2 (20 Hz) [20]. In another study rmse values of 8-10 cm were obtained when comparing SARAL data (40 Hz) with a tide gauge located in the Strait of Gibraltar along two $0 - 30$ km track segments [23]. In the German Bight and West Baltic Sea values of average stdd and correlation of 4.4 cm and 0.96 were obtained in the $0 - 10$ km segment [14]. In another study, the use of data from the same tide gauge at Huelva used in this study and CryoSat-2 data (20 Hz) in the $5 - 20$ km track segment yielded rmse values between 6.4-8.5 cm [10]. The validation of time series of Sea Surface Heights from SARAL (40 Hz), Jason-1/2, Envisat and ERS-2 (all at 20 Hz) was also performed along the French Atlantic coast in the southern Bay of Biscay [64]. The authors analysed the $0 - 5$ km coastal strip and found a rmse ranging between 8 cm (SARAL) and 89 cm (ERS-2). In the Gulf of Finland, a validation study of one year of S3A-SRAL Sea Surface Heights, yielded a rmse of 7 cm with respect to the tide gauges [65].

A deeper investigation of the $0 - 5$ km segment is presented here. The along-track values of rmse are shown in Figure 11 at the three locations. The two tracks at Huelva (Figure 11.a) show similar rmse values in the $3 - 5$ km segment than in the $5 - 20$ km segment (Figure 6a), and increase closer to land. This indicates that good quality data can be obtained within these tracks up to 3 km from the coast. Similarly, accurate S3A-SRAL data are observed at Barcelona over track #008 ($3 - 5$ km segment) (Figure 11.b). On the contrary, a high rmse is observed in the $0 - 5$ km segment at Barcelona and Bilbao for tracks #356 (Figure 11.b) and #051 (Figure 11.c), respectively. As previously mentioned, the $0 - 5$ km segment at Bilbao presented less than 20% of valid data for track #071 (Figure 4c). This evidences that a common track segment with the same level of accuracy cannot be achieved in our study areas. A possible explanation could be the type of transition of the tracks, i.e., ocean-to-land and land-to-ocean (Table I). However, this is not the case as good results are found in the land-to-ocean transitions at Huelva (#322) and Barcelona (#008) in the $3 - 5$ km segment. A second explanation might be the orientation of the coastline with respect to the track; this could be affecting the retracking of the waveforms due to the land contamination in the SAR altimetry footprint, giving inaccurate retracked *Ranges* and SWH. The smallest angle between each track and the coastline (θ) has been obtained (Table I). An angle of 90º implies that the track is perpendicular to the coast, so the land contamination on the radar waveforms might be neglected until very close to the coastline. Angles close to zero should complicate the retracking due to land contamination and hence, the accuracy of the sea level retrievals. Track #114 at Huelva is far from being perpendicular to the coast, with θ ~ 46º (Figure 1.a), but shows similar results in terms of rmse to track #322 (Figure 11.a), which is closer to perpendicular (θ ~ 75º). In the case of Barcelona, track #008 has a θ of ~ 61º and shows much more accurate results with respect to track #356 (Figure 11.b), with θ ~ 39º. Finally, Bilbao does not present accurate values for any of the two tracks in the $0 - 5$ km segment (θ = 74º and 88º for tracks #051 and #071, respectively). Hence, the angle between the tracks and the coastline does not explain the results in terms of rmse at Huelva and Bilbao. A third possibility is the land topography in the vicinity of the radar measurements footprint area. Figure 12 shows the $0 - 5$ km segment of the two tracks at Huelva: #114 (Figure 12.a) and #322 (Figure 12.b) and Bilbao: #051 (Figure 12.c) and #071 (Figure 12.d). Cycle 10 was used to show the track position. Also included is the land topography obtained from the SRTM data. The focus of this analysis is on the topography of the land located at a maximum perpendicular distance of 9.5 km of either sides of the track segment. This is the approximated beam-limited footprint of the S3A-SRAL radar altimeter (across-track direction). Considering the distance from the satellite's nadir to land and its height with respect to the sea level at nadir, one might expect the generation of simultaneous echoes with the water at nadir and hence the contamination of the waveforms. Also plotted in Figure 12 are the limits of the area inside the beam-limited footprint. Track #114 (Figure 12.a) shows land areas inside the footprint but the topography is very low (< 2 m) and almost flat, so no land contamination is expected in the vicinity of the leading edge of the waveforms corresponding to the $3 - 5$ km segment. Track #322 (Figure 12.b) does not show land inside the footprint due to its orientation with respect to the coast, so the small rmse in the $3 - 5$ km segment is well explained. The rmse values observed in the $0 - 3$ km segment might be due to the inaccuracy of some of the range and geophysical corrections applied to estimate S3A_SLA. In the case of Bilbao, track #051 (Figure 12.c) is affected by the proximity of Punta Galea in the eastern side of the track segment, which might justify a strong land contamination in the radar waveforms and hence, inaccurate retrievals of the geophysical parameters. This could explain the rmse observed along the whole $0 - 5$ km segment. Track #071 (Figure 12.d) shows land areas with a steep topography inside the footprint in the whole segment and thus, a strong land contamination is also expected.

## VII. CONCLUSION

This paper presents the results of the validation of six track segments of S3A-SRAL altimeter near the coasts of Spain. Three tide gauges located at three areas characterised by coastal environments with different tidal and hydrodynamic conditions were used for comparison: Huelva, Barcelona, and Bilbao. The number of valid data near the coast at the three locations, the along-track r and rmse of the $0 - 20$ km track segment of each track, the orientation of the satellite tracks respect to the land intersection, and the land contamination of the radar waveforms in the $0 - 5$ km track segments were analysed in detail. From the results obtained, the following conclusions are outlined.

The retrieval of the Range and SWH from the S3A-SRAL waveforms using the SAMOSA+ retracking algorithm and the use of the corrections (Range and Geophysical) available in the SARvatore GPOD service (with the exception of the SSB correction), give accurate along-track SLA up to very close to the shore (3 km in some cases) at the highest posting rate: 80 Hz. This accuracy was confirmed by comparing the altimeter SLA with the tide gauge data. To our knowledge, this is the first time that S3A-SRAL sea level data are validated at the highest posting rate available as a standard product. The tidal model available in GPOD (TPX08) shows a good performance in the three sites. The rss values are in line with other global tidal models: 1.97 cm (Huelva), 0.52 cm (Barcelona), and 5.45 cm (Bilbao). The SSB correction was not available at 80 Hz in GPOD at the time of downloading the data and, hence, a percentage of SWH was applied instead. The use of 5% of SWH as a first approximation seems to be the optimal choice at the three locations to get the best results in terms of smaller rmse and larger r.

Data screening plays a key role in the selection of the altimeter valid data. The comparison against tide gauges spotted anomalous sea level data not removed in the standard screening used in this work. Once these data were not considered in the comparison the rmse and r values improved by about 10%.

A common distance to the coast with a similar level of accuracy in the SLA from S3A-SRAL was not achieved in the three locations analysed. At Huelva (tracks #114 and #322) and Barcelona (track #008), accurate sea level data were observed in the 3 – 20 km track segment (r = 0.7-0.8 and rmse = 6-8 cm with more that 95% of valid data). The same level of accuracy was obtained at Barcelona (track #356) and Bilbao (tracks #051 and #071) in the 7 – 20 km track segment. The good results found at Huelva in the 3 – 5 km track segment are explained by the fact that the height of the land topography inside the beam-limited footprint of the altimeter is very close to zero and no bright targets on land seem to affect the retracking. Therefore, negligible land contamination is expected very close to the shore even if the track orientation with respect to the land is not optimal ($\theta = 48^\circ$ for track #114). However, the track orientation with respect to the land can explain the different behaviour found at Barcelona between track #008 ($\theta = 61^\circ$) and #356 ($\theta = 39^\circ$) because the topography in this region is steeper than in Huelva.

The abrupt topography observed at Bilbao around track #071 and the non-optimal orientation of track #051 explain the poor accuracy found in the 0 – 5 km track segment. Thus, knowledge of the track orientation with respect to the land and the land topography is needed to explain the results observed by satellite radar altimeters very close to the shore.

## ACKNOWLEDGMENTS

The altimetry data used in this study were obtained from the ESA-GPOD web page, and the tide gauge data were provided by the Spanish Puertos del Estado. Authors thank Salvatore Dinardo from HE Space for his technical support in the use of the GPOD service, and all the team members of GPOD. This article is included within a thesis supported by the PhD program of the University of Cadiz (Spain) and the University of Ferrara (Italy).

## REFERENCES

[1] Cazenave, A. (2018). Global sea-level budget 1993-present. Earth System Science Data, 10, 1551-1590.DOI: 10.5194/essd-10-1551-2018

[2] Vignudelli, S., Kostianoy, A. G., Cipollini, P., & Benveniste, J. (Eds.). (2011). Coastal altimetry. Springer Science & Business Media. e-ISBN: 978-3-642-12796-0.

[3] Gommenginger, C., Thibaut, P., Fenoglio-Marc, L., Quartly, G., Deng, X., Gómez-Enri, J., Challenor, P., Gao,Y. (2011). Retracking altimeter waveforms near the coasts. In Coastal altimetry (pp. 61-101). Springer, Berlin, Heidelberg.

[4] Brown, S. (2010). A novel near-land radiometer wet path-delay retrieval algorithm: Application to the Jason-2/OSTM Advanced Microwave Radiometer. IEEE Transactions on Geoscience and Remote Sensing, 48(4 PART 2), 1986–1992 DOI:10.1109/TGRS.2009.2037220.

[5] Brown, S. (2013). Maintaining the long-term calibration of the Jason-2/OSTM advanced microwave radiometer through intersatellite calibration. IEEE Transactions on Geoscience and Remote Sensing, 51(3), 1531–1543. DOI: 10.1109/TGRS.2012.2213262

[6] Fernandes, M. J., Lázaro, C., Nunes, A. L., Pires, N., Bastos, L., Mendes, V. B. (2010). GNSS-derived path delay: An approach to compute the wet tropospheric correction for coastal altimetry. IEEE Geoscience and Remote Sensing Letters, 7(3), 596–600. DOI: 10.1109/LGRS.2010.2042425

[7] Fernandes, M. J., Nunes, A. L., Lázaro, C. (2013). Analysis and inter-calibration of wet path delay datasets to compute the wet tropospheric correction for CryoSat-2 over ocean. Remote Sensing, 5(10), 4977–5005. DOI: 10.3390/rs5104977.

[8] Fenoglio-Marc, L., Dinardo, S., Scharroo, R., Roland, A., Dutour Sikiric, M., Lucas, B., Becker, M., Benveniste, J., Weiss, R., (2015). The German Bight: a validation of Cryosat-2 altimeter data in SAR mode. Adv. Space Res. 55, 2641–2656. DOI: 10.1016/j.asr.2015.02.014.

[9] Gómez-Enri, J., Scozzari, A., Soldovieri, F., Coca, J., Vignudelli, S. (2016). Detection and characterization of ship targets using CryoSat-2 altimeter waveforms. Remote Sensing, 8(3), 193. DOI: 10.3390/rs8030193.

[10] Gómez-Enri, J., Vignudelli, S., Cipollini, P., Coca, J., González, C. J. (2018). Validation of CryoSat-2 SIRAL sea level data in the eastern continental shelf of the Gulf of Cadiz (Spain). Advances in Space Research, 62(6), 1405-1420. DOI: 10.1016/j.asr.2017.10.042.

[11] Fernandes, M. J., Lázaro, C. (2016). GPD+wet tropospheric corrections for CryoSat-2 and GFO altimetry missions. Remote Sensing, 8(10), 1–30. DOI: 10.3390/rs8100851

[12] Passaro, M., Dinardo, S., Quartly, G.D., Snaith, H.N., Benveniste, J., Cipollini, P., Lucas, B. (2016). Cross-calibrating ALES Envisat and CryoSat-2 Delay-Doppler: a coastal altimetry study in the Indonesian Seas. Adv. Space Res. 58, 289–303. DOI: 10.1016/j.asr.2016.04.011.

[13] Passaro, M., Nadzir, Z. A., Quartly, G. D. (2018). Improving the precision of sea level data from satellite altimetry with high-frequency and regional sea state bias corrections. Remote Sensing of Environment, 218 (September), 245–254. DOI: 10.1016/j.rse.2018.09.007.

[14] Dinardo, S., Fenoglio-Marc, L., Buchhaupt, C., Becker, M., Scharroo, R., Joana Fernandes, M., Benveniste, J. (2018). Coastal SAR and PLRM altimetry in German Bight and West Baltic Sea. Advances in Space Research, 62(6), 1371–1404.DOI: 10.1016/j.asr.2017.12.018.

[15] Andersen, O.B. & Scharroo, R. (2011). Range and Geophysical Corrections in Coastal Regions: And Implications for Mean Sea Surface Determination. In Coastal altimetry (pp. 103-145). Springer, Berlin, Heidelberg.

[16] Tran, N, Vandemark, D, Labroue, S, Feng, H. Chapron, B., Tolman, H. L., Lambin, J., & Picot, N. (2010). Sea state bias in altimeter sea level estimates determined by combining wave model and satellite and satellite data. Journal of Geophysical Research, 115, 1–7. doi: 10.1029/2009JC005534.

[17] Pires, N., Fernandes, M., Gommenginger, C., & Scharroo, R. (2016). A conceptually simple modeling approach for Jason-1 sea state bias correction based on 3 parameters exclusively derived from altimetric information. Remote Sensing, 8(7), 576.

[18] Passaro, M., Nadzir, Z. A., & Quartly, G. (2018). Improving the precision of sea level data from satellite altimetry with high-frequency and regional sea state bias corrections. Remote Sensing of Environment, 245-254, doi: 10.1016/j.rse.2018.09.007.

[19] Cipollini, P., Benveniste, J., Birol, F., Fernandes, M. J., Obligis, E., Passaro, M. Strub, P. T., Valladeau, G., Vignudelli, S., Wilkin J. (2017). Satellite altimetry in coastal regions. In D. Stammer and A. Cazenave, (Eds.), Satellite Altimetry Over Oceans and Land Surfaces (pp.343-380), CRC Press. ISBN: 9781498743457.

[20] Birol, F., Delebecque, C. (2014). Using high sampling rate (10/20 Hz) altimeter data for the observation of coastal surface currents: A case study over the northwestern Mediterranean Sea. Journal of Marine Systems, 129, 318-333.

[21] Verron, J., Sengenes, P., Lambin, J., Noubel, J., Steunou, N., Guillot, A., Picot, N., Coutin-Faye, S., Sharma, R., Gairola, R.M., Murthy, R., Richaman, J., Griffin, D., Pascual, A., Rémy, F., Grupta, P.K. (2015). The SARAL/AltiKa altimetry satellite mission. Marine Geodesy, 38(sup1), 2-21. DOI: 10.1080/01490419.2014.1000471

[22] Bonnefond, P., Verron, J., Aublanc, J., Babu, K., Bergé-Nguyen, M., Cancet, M, Chaudhary, A., Crétaux, J., Frappart, F., Haines, B., Laurain, O, Ollivier, A., Poisson, J., Prandi, P., Sharma, R., Thibaut, P., Watson, C. (2018). The benefits of the Ka-Band as evidenced from the SARAL/AltiKa Altimetric mission: Quality assessment and unique characteristics of AltiKa data. Remote Sensing, 10(1), 83. DOI: 10.3390/rs10010083.

[23] Gómez-Enri, J., Cipollini, P., Passaro, M., Vignudelli, S., Tejedor, B., Coca, J. (2016). Coastal Altimetry products in the Strait of Gibraltar. IEEE Transactions on Geoscience and Remote Sensing, 54(9), 5455-5466. DOI: 10.1109/TGRS.2016.2565472.

[24] Gommenginger, C., Martin-Puig, C., Amarouche, L., Raney, K.R. (2014). Review of State of Knowledge for SAR Altimetry Over Ocean. Eumetsat Report. EUM/RSP/REP/14/749304.

[25] Peliz, A., Dubert, J., Marchesiello, P., Teles-Machado, A. (2007). Surface circulation in the Gulf of Cadiz: Model and mean flow structure. Journal of Geophysical Research-Oceans 112. DOI: 10.1029/2007JC004159.

[26] García-Lafuente, J., Delgado, J., Criado-Aldeanueva, F., Bruno, M., del Río, J., Vargas, J. M. (2006). Water mass circulation on the continental shelf of the Gulf of Cadiz. Deep Sea Research Part II: Topical Studies in Oceanography, 53(11-13), 1182-1197. DOI: 10.1016/j.dsr2.2006.04.011.

[27] Stevenson, R.E. (1977). Huelva Front and Malaga, Spain, Eddy chain as defined by satellite and oceanographic data. Deutsche Hydrographische Zeitschrift 30 (2), 51–53. DOI: 10.1007/BF02226082.

[28] Relvas, P., Barton, E. D. (2002). Mesoscale patterns in the Cape Sao Vicente (Iberian peninsula) upwelling region. Journal of Geophysical Research: Oceans, 107(C10), 28-1. DOI: 10.129/2000JC000456.

[29] Criado-Aldeanueva, F., García-Lafuente, J., Navarro, G., Ruiz, J. (2009). Seasonal and interannual variability of the surface circulation in the eastern Gulf of Cadiz (SW Iberia). Journal of Geophysical Research: Oceans, 114(C1). DOI: 10.1029/2008JC005069.

[30] Garel, E., Laiz, I., Drago, T., Relvas, P. (2016). Characterisation of coastal counter-currents on the inner shelf of the Gulf of Cadiz. Journal of Marine Systems, 155, 19-34. DOI: 10.1016/j.jmarsys.2015.11.001.

[31] Laiz, I., Gómez-Enri, J., Tejedor, B., Aboitiz, A., Villares, P. (2013). Seasonal sea level variations in the gulf of Cadiz continental shelf from in-situ measurements and satellite altimetry. Continental Shelf Research 53, 77-88. DOI: 10.1016/j.csr.2012.12.008

[32] Gómez-Enri, J., Aboitiz, A., Tejedor, B., Villares, P. (2012). Seasonal and interannual variability in the Gulf of Cadiz: Validation of gridded altimeter products. Estuarine, Coastal and Shelf Science 96, 114-121. DOI: 10.1016/j.ecss.2011.10.013.

[33] Font, J., Salat, J., Tintoré, J. (1988). Permanent features of the circulation in the Catalan Sea. Oceanologica Acta, Special Issue.

[34] Arnau, P.., Liquete, C., Canals, M. (2004). River mouth plume events and their dispersal in the Northwestern Mediterranean Sea. Oceanography- Washington DC- Oceanography Society-, 17, 22-31.

[35] Canals, M., P. Arnau, C. Liquete, S. Colas, Casamor, J.L. (2004). Catalogue and Data Set on River Systems from Mediterranean Watersheds of the Iberian Peninsula. Universitat de Barcelona, Barcelona, 220 pp.

[36] Mason, E., Coombs, S. H., Oliveira, P. B. (2005). An overview of the literature concerning the oceanography of the eastern North Atlantic region. Relatorios Cientificos e Tecnicos IPIMAR Serie Digital, 33, 59. ISSN: 1645-863X.

[37] Iribar, J., Ibáñez, M. (1979). Subdivisión de la zona intermareal de San Sebastián en función de los datos obtenidos con mareógrafo. Actas del I Simposio Ibérico de Estudios del Bentos Marino. San Sebastián. 2: 521-524.

[38] Piraud, I., Marseleix, P., Auclair, F. (2003). Tidal and thermohaline circulation in the Bay of Biscay. Geophys. Res. Abstracts, 5, 07058. ISBN 0–674.

[39] Charria, G., Lazure, P., Le Cann, B., Serpette, A., Reverdin, G., Louazel, S., Batifoulier, F., Fumas, F., Pichon, A., Morel, Y. (2013). Surface layer circulation derived from Lagrangian drifters in the Bay of Biscay. Journal of Marine Systems, 109, S60-S76. DOI: 10.1016/j.jmarsys.2011.09.015.

[40] OSPAR Commission. (2000). Region IV: Bay of Biscay and Iberian Coast. OSPAR Commission.

[41] Ferrer, L., Caballero, A. (2011). Eddies in the Bay of Biscay: A numerical approximation. Journal of Marine Systems, 87(2), 133-144. DOI: 10.1016/j.jmarsys.2011.03.008.

[42] Pingree, R.D., Sinha, B., Griffiths, C.R. (1999). Seasonality of the European slope current (Goban Spur) and ocean margin exchange. C. Shelf Res., 19, 929–975. DOI: 10.1016/S0278-4343(98)00116-2.

[43] Lavin, A., Valdes, L., Sanchez, F., Abaunza, P., Forest, A., Boucher, J, Jegou, A.M. (2004). The Bay of Biscay: the encountering of the Ocean and the Shelf (18b, E). The Global Coastal Ocean: Interdisciplinary Regional Studies and Syntheses. The Sea, 14, 933-1001. ISBN0–674-.

[44] Lazure, P., Garnier, V., Dumas, F., Herry, C., Chifflet, M. (2009). Development of a hydrodynamic model of the Bay of Biscay. Validation of hydrology. Continental Shelf Research, 29(8), 985-997. DOI: 10.1016/j.csr.2008.12.017.

[45] Froidefond, J.M., Jégou, A.M., Hermida, J., Lazure, P., Castaing, P. (1998). Variabilité du panache turbide de la Gironde par télédétection. Effets des facteurs climatiques. Oceanol. Acta, 21 (2), 191–207. DOI: 10.1016/S0399-1784(98)80008-X.

[46] Hermida, J., Lazure, P., Froidefond, J.M., Jégou, A.M., Castaing, P. (1998). La dispersion des apports de la Gironde sur le plateau continental. Données in situ, satellitales et numériques. Oceanol. Acta, 21(2), 209–221. DOI: 10.1016/S0399-1784(98)80009-1.

[47] Ferrer, L., Fontán, A., Mader, J., Chust, G., González, M., Valencia, V., Uriarte, Ad., Collins, M. B. (2009). Low-salinity plumes in the oceanic region of the Basque Country. Continental Shelf Research, 29(8), 970-984.

[48] Pingree, R. D., Le Cann, B. (1990). Structure, strength and seasonality of the slope currents in the Bay of Biscay region. Journal of the Marine Biological Association of the United Kingdom, 70(4), 857-885. DOI: 10.1017/S0025315400059117.

[49] Donlon, C., Berruti, B., Buongiorno, A., Ferreira, M. H., Féménias, P., Frerick, J., Goryl, P., Klein, U., Laur, H., Mavrocordatos, C., Nieke, J., Rebhan H., Seitz, B., Stroede, J., Sciarra, R. (2012). The global monitoring for environment and security (GMES) sentinel-3 mission. Remote Sensing of Environment, 120, 37-57. DOI: 10.1016/j.rse.2011.07.024.

[50] Ray, C., Martin-Puig, C., Clarizia, M.P., Ruffini, G., Dinardo, S., Gommenginger, C., Benveniste, J. (2015). SAR altimeter backscattered waveform model. IEEE Trans. Geosci. Remote Sens. 53, 911–919. DOI: 10.1109/TGRS.2014.2330423.

[51] Dinardo, S. (2013). Guidelines for the SAR (Delay-Doppler) L1b Processing, Paris, France, European Space Agency.

[52] Pérez, B., Álvarez Fanjul, E., Pérez, S., de Alfonso, M., Vela, J. (2013). Use of tide gauge data in operational oceanography and sea level hazard warning systems. Journal of Operational Oceanography, 6(2), 1-18.

[53] Puertos del Estado. Área de Medio Físico. (2012) Red Costera de Boyas. Informe de datos de la boya de Barcelona. Periodo: Dic.2011-Nov.2012. Available in: http://www.puertos.es/es-es/oceanografia/Paginas/portus

[54] Puertos del Estado. Área de Medio Físico. (2015) Red de boyas de aguas profundas. Informe de datos de la boya de Golfo de Cádiz. Periodo: Dic.2014-Nov.2015. Available in: http://www.puertos.es/es-es/oceanografia/Paginas/portus.aspx

[55] Puertos del Estado. Área de Medio Físico. (2015) Red de boyas de aguas profundas. Informe de datos de la boya de Bilbao. Periodo: Dic.2014-Nov.2015. Available in: http://www.puertos.es/es-es/oceanografia/Paginas/portus.aspx

[56] Jarvis A., Reuter, H.I., Nelson, A., Guevara, E. (2008). Hole-filled seamless SRTM data, International Centre for Tropical Agriculture (CIAT). Available in: http://srtm.csi.cgiar.org.

[57] Handoko, E., Fernandes, M. J., Lázaro, C. (2017). Assessment of Altimetric Range and Geophysical Corrections and Mean Sea Surface Models—Impacts on Sea Level Variability around the Indonesian Seas. Remote Sensing, 9(2), 102. doi:10.3390/rs9020102

[58] Egbert, Gary D., Svetlana Y. Erofeeva. (2002).Efficient inverse modeling of barotropic ocean tides. Journal of Atmospheric and Oceanic Technology 19.2. 183-204. DOI: 10.1175/1520-0426.

[59] Andersen, O.B., Knudsen, P. (2009). The DNSC08 mean sea surface and mean dynamic topography. J. Geophys. Res.-Oceans 114, 1–12. DOI: 10.1029/2008JC005179.

[60] Andersen, O.B., (2010). The DTU10 Gravity Field and Mean Sea Surface, Second International Symposium of the Gravity Field of the Earth (IGFS2), Fairbanks. Alaska.

[61] Pawlowicz, R., Beardsley, B., Lentz, S. (2002). Classical tidal harmonic analysis including error estimates in MATLAB using T_TIDE. Computers & Geosciences, 28(8), 929-937.

[62] Oreiro, F., D'Onofrio, E., Grismeyer, W., Fiore, M., Saraceno. M. (2014). Comparison of tide model outputs for the northern region of the Antarctic Peninsula using satellite altimeters and tide gauge data. Polar Sci. 8, 10–23. DOI: 10.1016/j.polar.2013.12.001.

[63] Cheng, Y., Andersen, O.B., (2011). Multimission empirical ocean tide modelling for shallow waters and polar seas. J. Geophys. Res. 116 (C11001), 11. DOI: 10.1029/2011JC007172.

[64] Vu, P., Frappart, F., Darrozes, J., Marieu, V., Blarel, F., Ramillien, G., Bonnefond, P, Birol, F. (2018). Multi-satellite altimeter validation along the french atlantic coast in the southern bay of biscay from ers-2 to saral. *Remote Sensing*, *10*(1), 93.

[65] ] Birgiel, E., Ellmann, A., Delpeche-Ellmann, N. (2019). Performance of Sentinel-3A SAR Altimetry Retrackers: The SAMOSA Coastal Sea Surface Heights for the Baltic Sea.

BIOGRAPHY AND PHOTOS

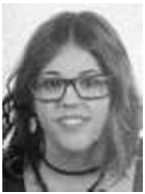

**Ana Aldarias** received the degree in Marine Sciences from the University of Cadiz, Cadiz, Spain, in 2015 and the MsC degree in Oceanography from the University of Cadiz, Cadiz, Spain, in 2016. In 2017, she did a predoctoral internship in the Environmental Fluid Dynamic research group of the Andalusian Institute of Earth System from the University of Granada, Granada, Spain. In 2018, she started to work in the Physical Oceanography and Remote Sensing research group of the Applied Physics department from the University of Cadiz, Cadiz, Spain. Last November, she got a grant from the PhD program –Earth and Marine Science‖, of the University of Cadiz (Spain) and the University of Ferrara (Italy). At present, she is doing the PhD thesis in physical oceanography and remote sensing.

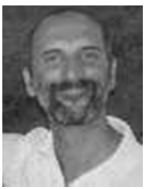

**Jesús Gómez-Enri** received the degree in Marine Sciences from the University of Cadiz, Cadiz, Spain, in 1996 and the Ph.D. degree in satellite Oceanography from the University of Cadiz, in 2002. From 1999 to 2001, he was a trainee at the European Space Agency European Space Research Institute (ESA-ESRIN), Frascati, Rome, Italy; from 2003 to 2005 he was a post-doctoral fellow at the National Oceanography Centre, Southampton, U.K. Since 2005, he has been an Associate Professor with the Applied Physics Department, University of Cadiz. His teaching and research is focused on Remote Sensing applied to the Ocean. He works in satellite radar altimetry and synthetic aperture radar.

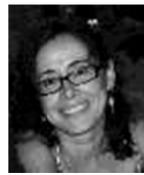

**Irene Laiz** undertook her undergraduate studies at the University of Las Palmas de Gran Canaria (ULPGC), Spain, where she graduated with 1st Class Hons in Marine Sciences in 1997. She received her Ph.D. degree in Physical Oceanography in 2005 from the ULPGC. Between 2001-2006, she worked as a Research Scientist at BMT Cordah Ltd. (Southampton, U.K.), specialising in the development and implementation of numerical models dealing with marine pollution. Between 2006-2010, she worked as postdoctoral fellow at the Andalusian Institute of Marine Sciences (ICMAN-CSIC, Cadiz, Spain). She joined the University of Cadiz Applied Physics Department as a postdoctoral fellow in 2011, and remained there as a lecturer since 2013. Her research is mainly focused on ocean circulation and satellite altimetry.

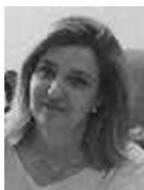

**Begoña Tejedor** received her degree in Physics from the Compluntense University of Madrid, Spain, in 1987; specialist in "Physics of the Earth, Air & Cosmos". She received her Ph.D. degree in 1991 from Las Palmas de Gran Canaria University (LPGCU). In 1987, she joined as Associate Professor at the Applied Physics Department (Faculty of Marine Sciences), at the LPGCU, and later on as Associate Professor and Head of the Applied Physics Department in the Faculty of Marine Sciences at the University of Cadiz (1991). Her teaching and research has been focused on Physical Oceanography. One of her line of research is the study of physical processes in shallow waters using in-situ and Remote Sensing data.

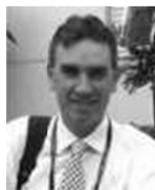

**Stefano Vignudelli** received the doctoral degree in Engineering from University of Pisa, Italy, in 1991. He is currently a researcher employed at the Consiglio Nazionale delle Ricerche (National Research Council) in Pisa, Italy. He has over 20 years of scientific experience in the area of satellite Remote Sensing (radar altimetry, in particular) for studying coastal and marine/inland environments (water level variability, in particular). Major interests include processing methods for data analysis, validation with local field observations, multi-sensor synergy, and exploitation. Most significant accomplishment has been to lead development of satellite radar altimetry in challenging areas (e.g., coastal zone) to provide improved measurements for water level research and applications.

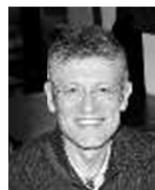

**Paolo Cipollini** (S'93–M'97–SM'03) received the laurea (M.Eng.) degree in electronics engineering from the University of Pisa, Pisa, Italy, in 1992 and the Ph.D. degree in methods and technologies for environmental monitoring from the University of Florence, Firenze, Italy, in 1996. Paolo is an enginer ad satellite oceanographer with over 20 years' experience. From 1997 to 2018 he was at the National Oceanography Centre in Southampton, U.K., where he worked on the detection and characterization of oceanic planetary waves (Rossby waves) and on development and application of satellite radar altimetry in the coastal zone, also managing the ESA COASTALT project (2008–2012) for the development of coastal altimetry for Envisat and being a Principal Investigator in Cryosat-2 and Sentinel-3 projects. Since 2018 he supports the activities of the European Space Agency Climate Office in Harwell, U.K., in particular the exploitation of ocean remote sensing for climate research within the ESA Climate Change Initiative (CCI) Program, interacting directly with scientists all over Europe, the user communities and with the international partner organisations of ESA.

TABLES CAPTION

Table I. Information about the number of track, location, latitude and longitude of the tide gauges (TG), minimum distance between each tide gauge and the tracks of S3A-SRAL in this zone, transition: ocean to land (OL) or land to ocean (LO) and the smaller angle of the track respect to the coast.

Table II. Amplitudes (cm) and phases (°) of the main tidal constituents in Huelva, Barcelona, and Bilbao. Data from the Puertos del Estado tide gauges (TG) and the global tidal model TPX08.

Table III: Assessment of the TPX08 model in the closest point to the tide gauges. Comparison of amplitude and phase of the main tidal constituents using the statistics: root mean squares (rms) and residual sum of squares (rss).

Table IV. Results of the comparison between the altimeter and tide gauge SLA time series in the study areas (HU: Huelva, BA: Barcelona, BI: Bilbao). Average of rmse (cm), r, confidence level (CL) and percentage of valid data (VD) for the different length of the track segments: 0-5 km, 5-20 km and 0-20 km respect to the coast.

FIGURES CAPTION

Figure 1. Study areas: (a) Huelva, (b) Barcelona, (c) Bilbao, and their location in the Iberian Peninsula (d). Track segments of the Sentinel-3A orbits analysed: Black-line: 5-20 km segment; red line: 0-5 km segment. Blue-star: tide gauge positions. Blue-point: closest points of the tidal model to the tide gauges.

Figure 2. Wave roses for 2016-2018 period from (a) Gulf of Cadiz, (b) Barcelona, and (c) Bilbao-Vizcaya buoys.

Figure 3. Rmse (cm) between S3A_SLA and TG_SLA time series as a function of the % of SWH used for the SSB correction at (a) Huelva, (b) Barcelona, and (c) Bilbao. Blue line: Ocean to land transition. Red line: Land to Ocean transition.

Figure 4. Number of valid data in the track segment 0 - 20 km for (a) Huelva, (b) Barcelona, and (c) Bilbao. The dashed black line (vertical) indicates the position of the 5 km distance to the coast. The dashed black line (horizontal) gives the limit of 20% of valid data.

Figure 5. Along-track values of r in the $5 - 20$ km track segment at (a) Huelva, (b) Barcelona, and (c) Bilbao.

Figure 6. Same as Figure 5 for rmse.

Figure 7. (a) Cycle-by-cycle S3A_SLA values in the $5 - 20$ track segment for track #071. (b) S3A_SLA from cycle 032 (05/06/2018).

Figure 8. (a) Along-track SWH of track #071 (cycle 032) (Bilbao) along the $5 - 20$ km track segment. (b) Radargram of the waveforms (only from gate 300 to 400) (power in watts). The positions corresponding to 12 and 17 km to the coast are indicated in black (8.a) and white (8.b) dashed lines.

Figure 9. Along-track values of (a) r and (b) rmse of track #071 considering all the cycles after data screening (blue line), and removing cycle 32 (red line).

Figure 10. Same as Figure 9 for track #008 including all the cycles after data screening (blue line), and removing cycle 06 (red line).

Figure 11. Same as Figure 6 for the $0 - 5$ km track segment at (a) Huelva, (b) Barcelona, and (c) Bilbao. Missing track #071 due to data unavailability.

Figure 12. Location of the $0 - 5$ km track segments at Huelva for tracks #114 (a) and #322 (b), and at Bilbao for tracks #051 (c) and #071 (1d). The SRTM land topography (in m) is also shown. The envelope of the beam-limited footprint in the across-track direction (a radius of about 9.5 km perpendicular to the track) is delimited with a dashed red line.

Table I: Information about the number of track, location, latitude and longitude of the tide gauges (TG), minimum distance between each tide gauge and the tracks of S3A-SRAL in this zone, transition: ocean to land (OL) or land to ocean (LO) and the smaller angle of the track respect to the coast.

| Track nb. | #114 | #322 | #356 | #008 | #051 | #071 |
|---|---|---|---|---|---|---|
| Location | Huelva | | Barcelona | | Bilbao | |
| TG position | 37.13º N - 6.83º W | | 41.34º N - 2.17º E | | 43.35º N - 3.05º W | |
| Min. distance TG - S3A-SRAL | 16.2 km | 14.9 km | 4.0 km | 32.4 km | 0.9 km | 7.7 km |
| Type of transition | OL | LO | OL | LO | OL | LO |
| Track direction | Ascending | Descending | Ascending | Descending | Descending | Ascending |
| Time of over passing | 22:00 | 10:54 | 21:21 | 10:23 | 10:45 | 21:40 |
| Angle respect to the coast | 46º | 75º | 39º | 61º | 74º | 88º |

Table II: Amplitudes (cm) and phases (º) of the main tidal constituents in Huelva, Barcelona, and Bilbao. Data from the Puertos del Estado tide gauges (TG) and the global tidal model TPX08.

| | Tidal constituents: | | M2 | S2 | N2 | K2 | K1 | O1 | Q1 | P1 | M4 |
|---|---|---|---|---|---|---|---|---|---|---|---|---|
| HU | Amp (cm) | TG | 104.0 | 37.8 | 22.2 | 10.6 | 6.6 | 5.8 | 1.7 | 2.3 | 2.8 |
| | | TPX08 | 102.1 | 37.2 | 21.7 | 10.3 | 6.6 | 6.0 | 1.8 | 2.0 | 1.7 |
| | Pha (º) | TG | 57.4 | 84.1 | 41.3 | 81.1 | 48.1 | 311.1 | 228.0 | 41.9 | 170.2 |
| | | TPX08 | 57.1 | 83.1 | 41.1 | 79.3 | 50.5 | 308.2 | 257.5 | 45.0 | 156.4 |
| BA | Amp (cm) | TG | 4.6 | 1.7 | 1.0 | 0.5 | 3.7 | 2.4 | 0.3 | 1.2 | 0.5 |
| | | TPX08 | 4.5 | 1.5 | 1.0 | 0.5 | 3.6 | 2.2 | 0.3 | 1.2 | 0.36 |
| | Pha (º) | TG | 213.5 | 230.7 | 201.5 | 228.6 | 167.8 | 102.9 | 51.3 | 160.4 | 346.8 |
| | | TPX08 | 220.5 | 238.4 | 204.0 | 233.1 | 167.1 | 108.4 | 71.56 | 155.55 | 356.3 |
| BI | Amp (cm) | TG | 131.1 | 45.9 | 27.8 | 13.0 | 6.4 | 7.0 | 2.2 | 2.0 | 2.7 |
| | | TPX08 | 133.7 | 46.5 | 27.9 | 13.1 | 6.6 | 6.6 | 2.2 | 2.1 | 2.5 |
| | Pha (º) | TG | 92.4 | 124.9 | 72.8 | 121.8 | 68.9 | 321.3 | 274.2 | 57.0 | 324.4 |
| | | TPX08 | 95.3 | 127.2 | 85.7 | 125.0 | 74.3 | 323.6 | 275.2 | 64.6 | 306.9 |

Table III: Assessment of the TPX08 model in the closest point to the tide gauges. Comparison of amplitude and phase of the main tidal constituents using the statistics: root mean squares (rms) and residual sum of squares (rss).

| MAIN TIDAL CONSTITUENTS | rms (cm) | | |
|---|---|---|---|
| | HUELVA | BARCELONA | BILBAO |
| M2 | 1.41 | 0.40 | 4.99 |
| S2 | 0.65 | 0.17 | 1.69 |
| N2 | 0.29 | 0.03 | 0.98 |
| K2 | 0.30 | 0.03 | 0.53 |
| K1 | 0.20 | 0.08 | 0.48 |
| O1 | 0.23 | 0.19 | 0.36 |
| Q1 | 0.65 | 0.08 | 0.03 |
| P1 | 0.23 | 0.08 | 0.20 |
| M4 | 0.86 | 0.16 | 0.58 |
| rss (cm) | 1.97 | 0.52 | 5.45 |

Table IV: Results of the comparison between the altimeter and tide gauge SLA time series in the study areas (HU: Huelva, BA: Barcelona, BI: Bilbao). Average of rmse (cm), r, confidence level (CL) and percentage of valid data (VD) for different portions of the track segments: 0-5 km, 5-20 km and 0-20 km respect to the coast.

| Zone | Track | 0 - 5 km | | | | 5 – 20 km | | | | 0 – 20 km | | | |
|---|---|---|---|---|---|---|---|---|---|---|---|---|---|
| | | rmse | r | CL | VD (%) | rmse | r | CL | VD (%) | rmse | r | CL | VD (%) |
| HU | #114 | 14.2 | 0.6 | 95.5 | 86.0 | 6.4 | 0.8 | 100 | 90.6 | 8.0 | 0.8 | 99.1 | 89.9 |
| | #322 | 16.4 | 0.6 | 92.1 | 87.9 | 6.0 | 0.8 | 100 | 98.8 | 8.6 | 0.7 | 98.0 | 96.1 |
| BA | #356 | 29.0 | 0.2 | 67.0 | 68.2 | 6.1 | 0.8 | 100 | 97.6 | 11.5 | 0.6 | 92.6 | 90.8 |
| | #008 | 12.3 | 0.7 | 98.5 | 85.7 | 6.0 | 0.8 | 100 | 99.1 | 7.6 | 0.8 | 99.6 | 95.8 |
| BI | #051 | 20.2 | 0.5 | 90.5 | 89.5 | 7.8 | 0.8 | 99.9 | 94.1 | 10.9 | 0.7 | 97.6 | 93.0 |
| | #071 | --- | --- | --- | 15.6 | 8.2 | 0.7 | 99.8 | 88.5 | 9.6 | 0.7 | 93.5 | 88.6 |

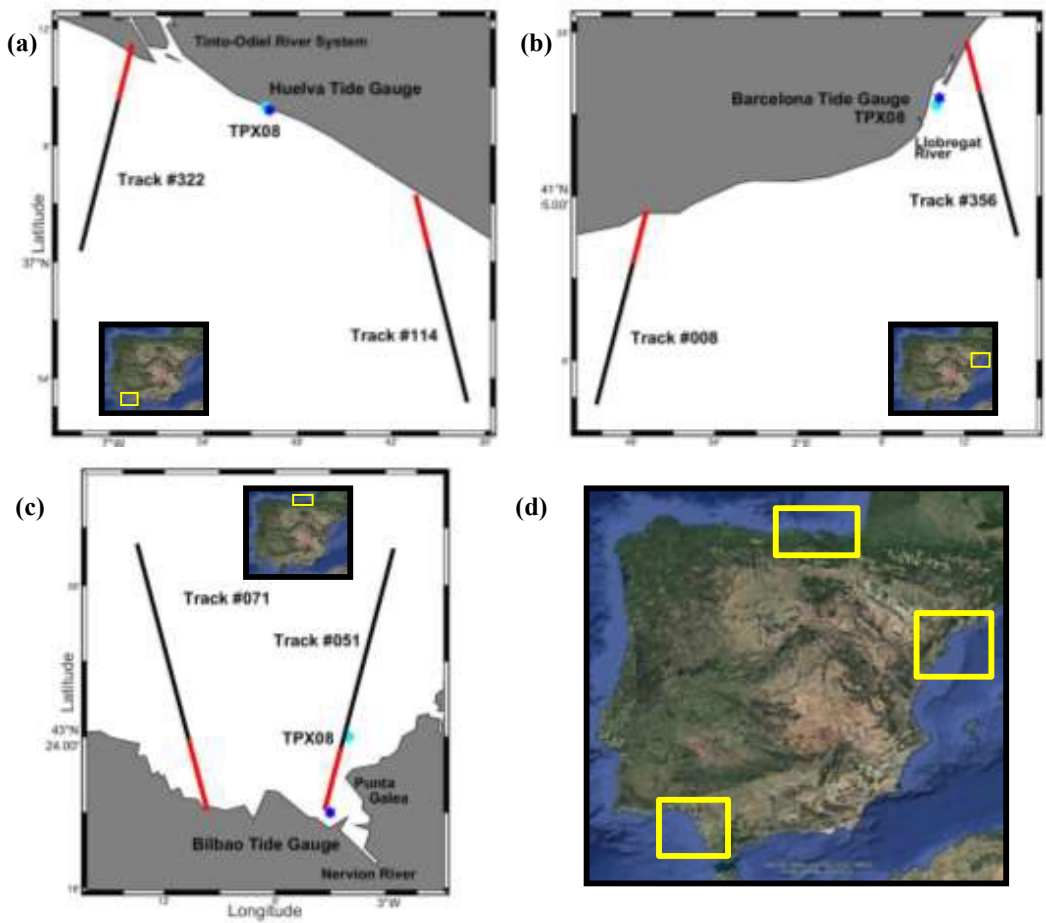

Figure 1. Study areas: (a) Huelva, (b) Barcelona, (c) Bilbao, and their location in the Iberian Peninsula (d). Track segments of the Sentinel-3A orbits analysed: Black-line: 5-20 km segment; red line: 0-5 km segment. Blue-star: tide gauge positions. Blue-point: closest points of the tidal model to the tide gauges.

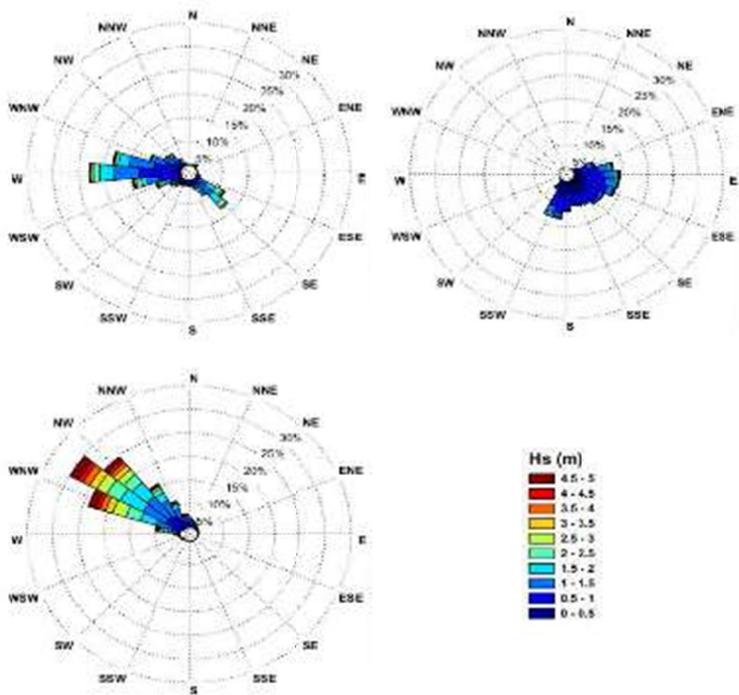

Figure 2. Wave roses for 2016-2018 period from (a) Gulf of Cadiz, (b) Barcelona, and (c) Bilbao-Vizcaya buoys.

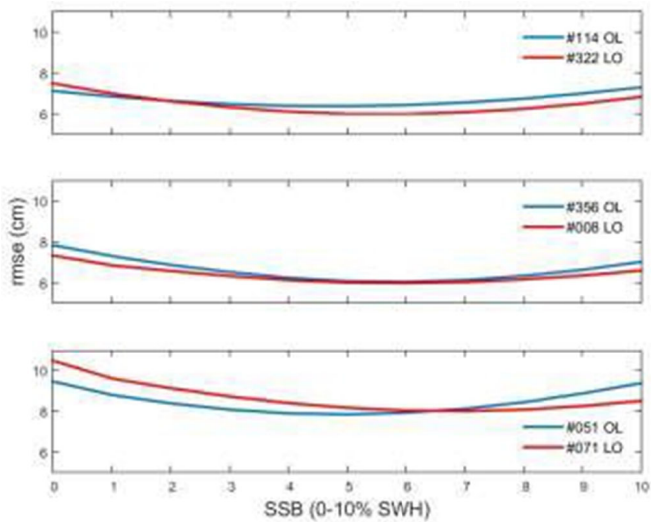

Figure 3. Rmse (cm) between S3A_SLA and TG_SLA time series as a function of the % of SWH used for the SSB correction at (a) Huelva, (b) Barcelona, and (c) Bilbao. Blue line: Ocean to land transition. Red line: Land to Ocean transition.

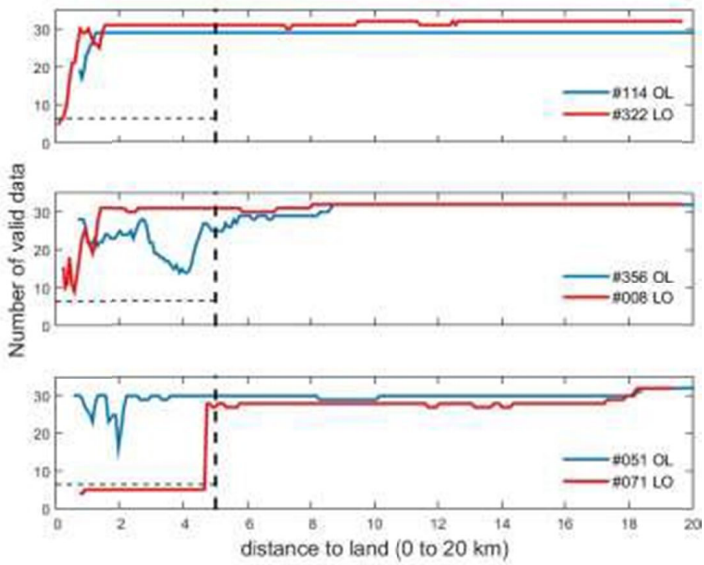

Figure 4. Number of valid data in the track segment 0 - 20 km (distance from the intersection of the track with the closest land) for (a) Huelva, (b) Barcelona, and (c) Bilbao. The dashed black line (vertical) indicates the position of the 5 km distance to the intersection with the coast. The dashed black line (horizontal) gives the limit of 20% of valid data.

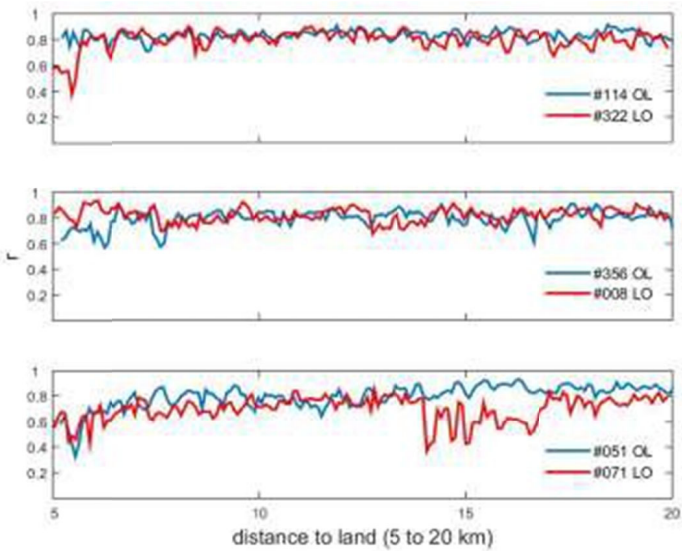

Figure 5. Along-track values of r in the 5 – 20 km track segment at (a) Huelva, (b) Barcelona, and (c) Bilbao.

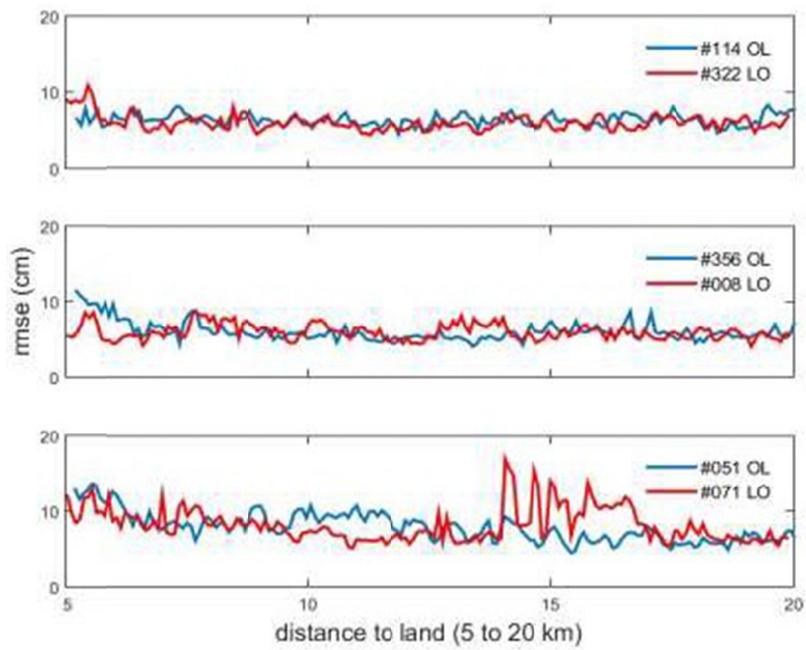

Figure 6. Same as Figure 5 for rmse.

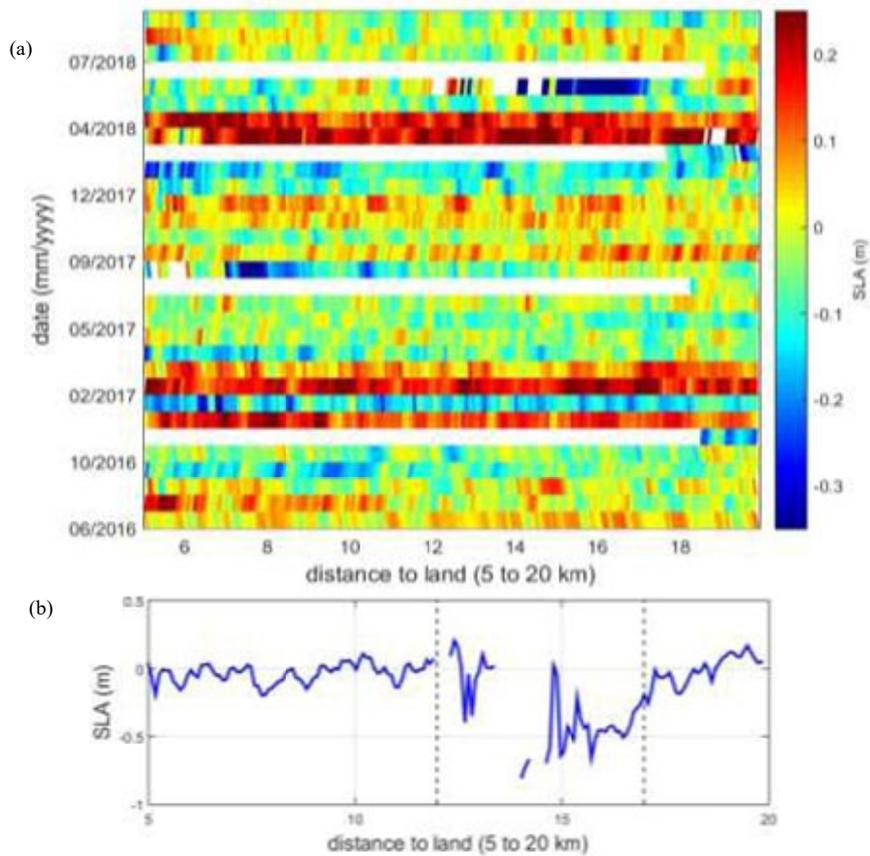

Figure 7. (a) Cycle-by-cycle S3A_SLA values in the 5 – 20 track segment for track #071. (b) S3A_SLA from cycle 032 (05/06/2018).

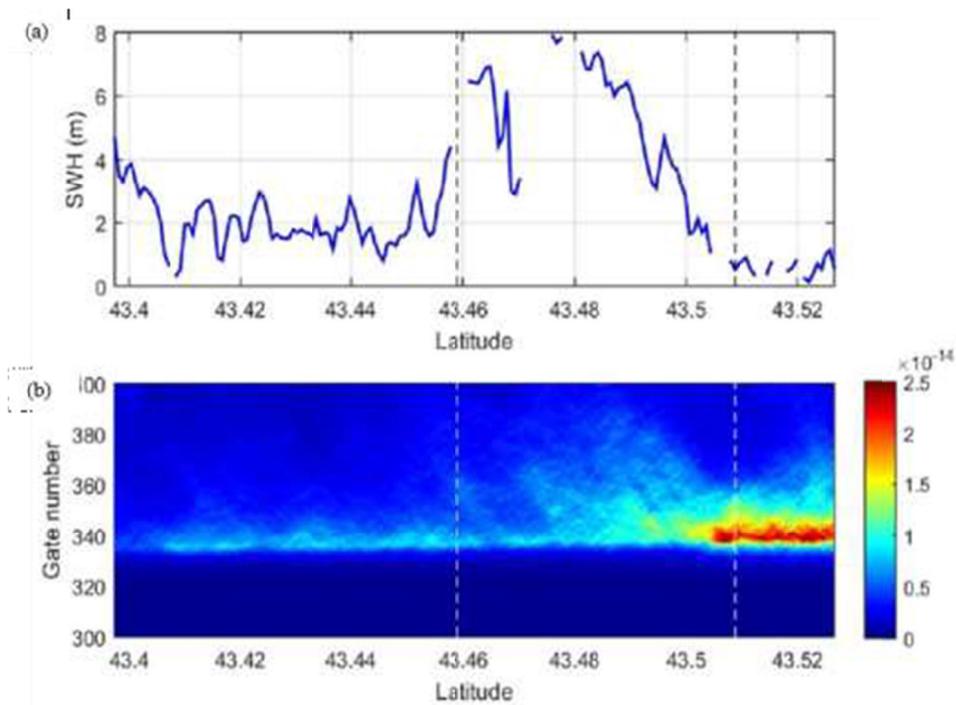

Figure 8. (a) Along-track SWH of track #071 (cycle 032) (Bilbao) along the 5 – 20 km track segment. (b) Radargram of the waveforms (only from gate 300 to 400) (power in watts). The positions corresponding to 12 and 17 km to the coast are indicated in black (8.a) and white (8.b) dashed lines.

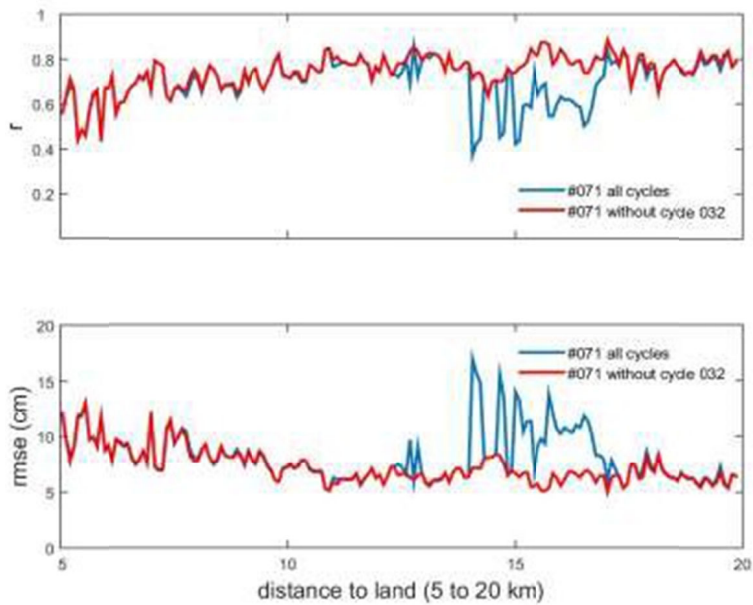

Figure 9. Along-track values of (a) r and (b) rmse of track #071 considering all the cycles after data screening (blue line), and removing cycle 32 (red line)

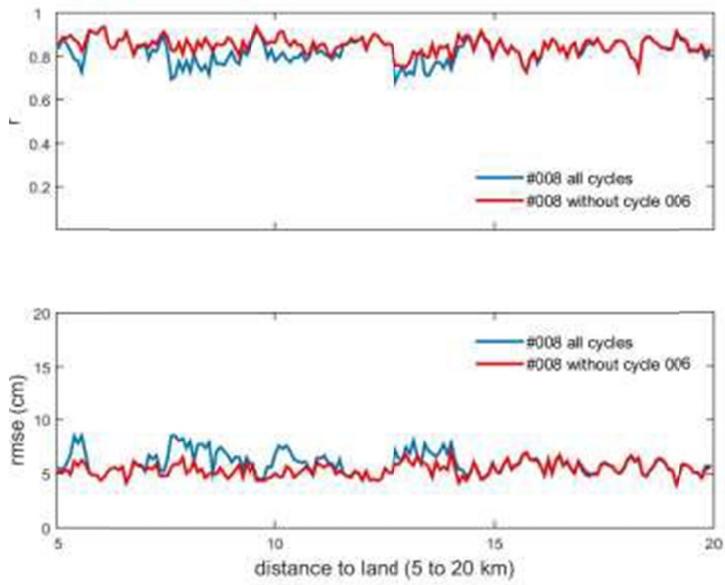

Figure 10. Same as Figure 9 for track #008 including all the cycles after data screening (blue line), and removing cycle 06 (red line)

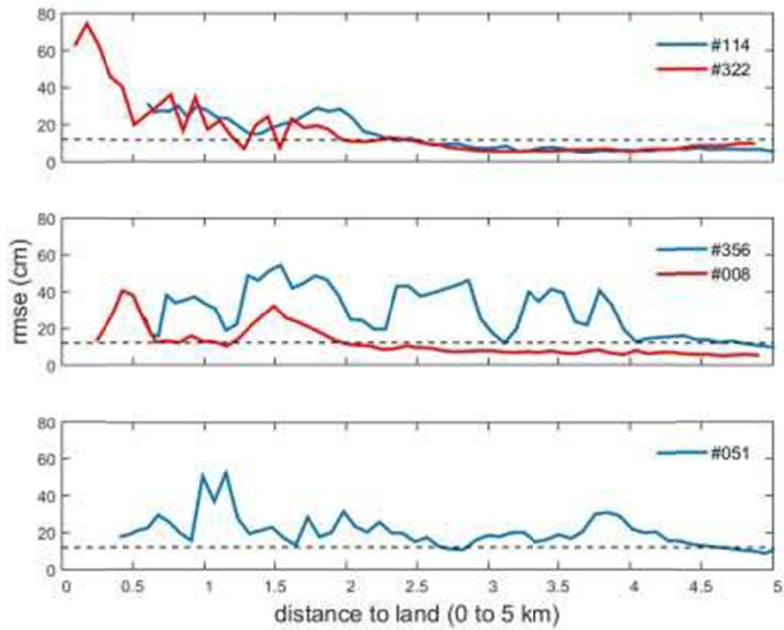

Figure 11. Same as Figure 6 for the 0 – 5 km track segment at (a) Huelva, (b) Barcelona, and (c) Bilbao. Missing track #071 due to data unavailability

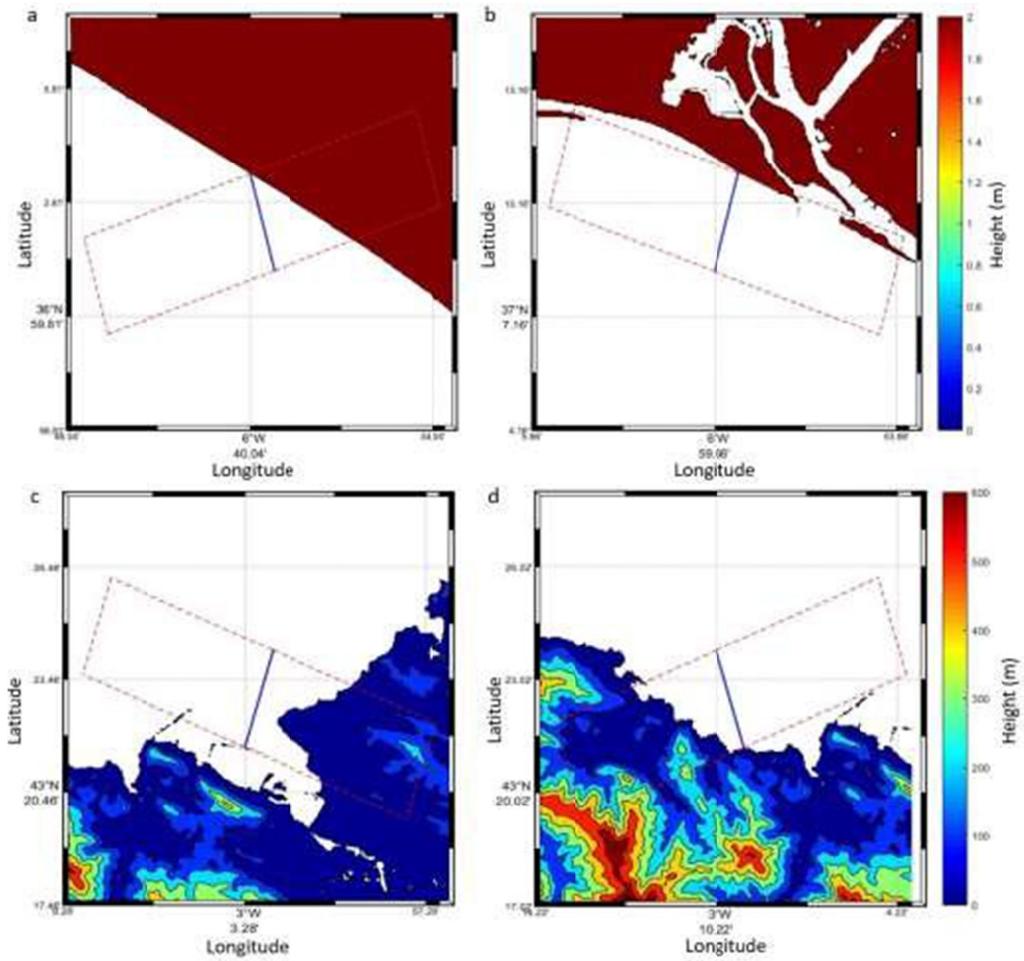

Figure 12. Location of the 0 – 5 km track segments at Huelva for tracks #114 (a) and #322 (b), and at Bilbao for tracks #051 (c) and #071 (1d). The SRTM land topography (in m) is also shown. The envelope of the beam-limited footprint in the across-track direction (a radius of about 9.5 km perpendicular to the track) is delimited with a dashed red line.